\documentclass{article}
\pdfoutput=1
\usepackage{ijcai23}

\usepackage{times}
\usepackage{soul}
\usepackage{url}
\usepackage[hidelinks]{hyperref}
\usepackage[utf8]{inputenc}
\usepackage[small]{caption}
\usepackage{graphicx}
\usepackage{amsmath}
\usepackage{amsthm}
\usepackage{booktabs}
\usepackage{algorithm}
\usepackage{algorithmic}

\usepackage{color}              
\usepackage{glossaries}         
\usepackage[inline]{enumitem}   
\usepackage{svg}                

\newacronym{amss}{AMSS}{Angular Metric for Shape Similarity}
\newacronym{dtw}{DTW}{Dynamic Time Warping}
\newacronym{ddtw}{DDTW}{Derivative Dynamic Time Warping}
\newacronym{farm}{FARM}{Forward Angular Relevance Metric}
\newacronym{rms}{RMS}{Root Mean Square}

\title{Exogenous Data in Forecasting: FARM - A New Measure for Relevance Evaluation}

\author{
 Ramón Christen$^{1,2}\dagger$
 \and
 Luca Mazzola$^1$\and
 Alexander Denzler$^1$\And
 Edy Portmann$^2$
 \affiliations
 $^1$School of Computer Sciences - Lucerne University of Applied Sciences and Arts\\
 $^2$Human-IST Institute - University of Fribourg
 \emails
 \textit{firstname}.\textit{lastname}@hslu.ch ,
 edy.portmann@unifr.ch\\
 $\dagger$ corresponding author: \textit{ramon.christen@hslu.ch}
}

\begin{document}
\maketitle

\begin{abstract}
Evaluating the relevance of an exogenous data series is the first step in improving the prediction capabilities of a forecast algorithm. Inspired by existing metrics for time series similarity, we introduce a new approach named FARM - Forward Aligned Relevance Metric. Our forward method relies on an angular measure that compares changes in subsequent data points to align time-warped series in an efficient way. The proposed algorithm combines local and global measures to provide a balanced relevance metric. This results in considering also partial, intermediate matches as relevant indicators for exogenous data series significance. As a first validation step, we present the application of our FARM approach to synthetic but representative signals. While demonstrating the improved capabilities with respect to existing approaches, we also discuss existing constraints and limitations of our idea.

    %
    %
\end{abstract}

\section{Introduction}
Time series usually rely on different external variables. Together, they build a system that controls the recorded, independent variable. Some of the external variables also depend on system's state. Others only affect the system but do not depend on it. 
In literature these variables are called exogenous variables.

Because of causality, these data might comprise important information about the behaviour of target variables for some time intervals. As result, specific information from exogenous variables may be used to infer future behaviour of target variables. 
In this sense, certain information can be valuable for increasing the forecast accuracy.

The potential of including exogenous variables in forecast was demonstrated in several experiments such as in \cite{hernandez_study_2012,de_felice_electricity_2013,tascikaraoglu_short-term_2016,zhu_short-term_2018} for energy forecasts. It shows, including exogenous data in consideration allows for more accurate forecast results. This is a key component for complex solutions in smart environments such as the management for a balanced local energy consumption and production. In addition, increasing data recordings from tracking and monitoring devices or social media and consumer applications provide new sources of potentially valuable exogenous data.


The impact of exogenous variables on the target variable differs - between variables but also the state of the exogenous variable itself.
While some have a strong continuous impact on the change of the target variable other only have a weak or very specific impact. For this reason, considering information provided by exogenous variables for forecast requires to know to what extent an exogenous variable affect the target time series' behaviour. This holds for the variable in general but also for intermediate behaviour.

State-of-the-art approaches use standard similarity measures such as the Mutual Information, Auto- or Pearson correlation in order to evaluate the relevance of variables such as in \cite{silva_impact_2019,huang_short_2016}. This can be a reasonable measure for variables on a common time base affected by the same environment. Yet, evaluating the impact of exogenous variable on a target series has twofold challenges: (1) time and amplitude warping between exogenous and target variable and (2) locate and credit influential attributes with a limited time span. Both are neither compensated nor represented in the output of standard similarity measures. In fact, such measures are not applicable to quantify the relevance of exogenous data on a target time series.


Indeed, literature presents different methodologies being capable to deal with warped time series. In a previous literature review the authors analysed and contrasted several approaches for comparing series with time and amplitude warping and quantifying the relevance 
\cite{christen_distance_2022}. Yet, many of these approaches are measures due to the lack in fulfilling the triangle inequality that is mandatory for a metric \cite{jakel_similarity_2008}. Secondly, each result in a single similarity and distance measure, respectively, without information of intermediate feature matches.

The review presented that approaches using the change between data points as measure for distance/similarity between time series which provide promising results for time warping compensation. This holds for
the shape-based similarity measure called \gls{amss} proposed in \cite{nakamura_shape-based_2013} and the \gls{ddtw} proposed in \cite{keogh_derivative_2001} among others. While the \gls{amss} measure relies
on vector sequence equivalently representing time series data, the \gls{ddtw} uses a linear combination of value differences between three subsequent data points.
As result, we worked on a new metric that relies on the concept of the \gls{amss} approach and additionally provides a series comprising location related distance information.

In this paper we introduce a new shape-based relevance metric called \gls{farm}
\footnote{The code will be made available after acceptance.}.
It relies on an angular distance measure that results from the difference of subsequent data points and that fulfils the triangle inequality. In addition, it implements a feed forward approach for feature matching and warping compensation that reduces algorithm's complexity compared to other approaches. Finally, it results in a similarity value that rely on intermediate and global measurements. 


The remainder of this paper is structured as follows: Section \ref{sec:asp_rel_metric} discusses the different aspects of a relevance metric that have to be treated separately for developing a new metric before we introduce the new \gls{farm} metric in Section \ref{sec:farm}. Subsequently, we present first validation results and reason constraints of the proposed metric in Section \ref{sec:lim_and_compl} before discussing further research and the conclusions in Sections \ref{sec:further_res} and \ref{sec:conclusion} respectively.

\section{Aspects of a Relevance Metric}
\label{sec:asp_rel_metric}
A relevance metric provides a measure for quantifying the information value (i.e. the relevance) provided by any variable with respect to a target time series. This holds for any pair of time series; such with a clear casual relationship as well as for time series without an obvious causality at a first glance.
Achieving reasonable measures from time series comparison requires identical prerequisites in acquiring technology and system dependency.
These differ depending on the measure used for comparing data points and the concept of the algorithm itself. As result, the algorithm used for quantifying the relevance of a time series has to treat thee key aspects:
\begin{enumerate*}[label={\roman*)}]
    \item distance/similarity measure,
    \item warping and scaling effects between compared time series and
    \item the quantification of features similarities.
\end{enumerate*}
Although there is no strict separation possible between these aspects, the analysis of existing and the development of a new metric should treat them separately. 
Following, we discuss the aspect in more detail towards the development of a new relevance metric.

\begin{description}[style=unboxed,leftmargin=0cm]
    \item[The distance/similarity measure] is the foundation for all subsequent decisions and actions. For this reason, it is mandatory having a reasonable measure that matches the final objectives. In particular, the measure has to reflect the information required for a suitable comparison. Basically, the distance/similarity consist of two parts: (1) a feature representing characteristics of time series and (2) a measure that quantifies the feature in order to use it for comparison and other actions.
    
    As feature (1), similarity measure often use the amplitude of time series; in particular the value of data points as in the \gls{dtw} algorithm for treating time warping issues. Alternatives range from amplitude derivation, coefficients of frequency or other spectrum, shape representation or statistic numbers to complex feature vectors composing multiple characteristics such as in \cite{chen_spade_2007}. The feature selection primarily depends on the requirements of subsequent actions. In other words, what the feature is used for.
    From this perspective, the feature representing specific attributes from time series has to correspond with the information required to treat the time series accordingly. This information vary for classification or forecasting purposes and is defined by the final objectives. 
    In the proposed algorithm, this requirement primarily focus on shape information under consideration of possible warping effects.

    The measure (2) quantifies the information represented in feature and makes the information available for numerical calculation. That means, the measure of a feature allows for defining a distance or similarity between individual attributes of compared time series. The resulting numerical (or categorical at least) value representing specific attribute information can then be used to identify matches between compared time series - for full time series or segments only. In case of feature vectors, either features will be treated differently or they result in one common measure.
    However, for using a measure as metric, it has to fulfil the triangular inequality \cite{jakel_similarity_2008}.
    
    Certainly, in combination with the feature selection, the measure will have a significant influence on the robustness of the final algorithm. For this reason, it is of utmost importance to define the features and the measure as elementary as possible for best robustness, yet as specific as required for to be able to extract enough information.
    
    \item[Warping and scaling effects] are common aspects in data comparison. Time warping can result from independent acquisition, varying response time of an impact or if data have no primary relationship (i.e. no causation) 
    to each other. In time series, this results in misaligned common features and causes bad results in a
    point-by-point
    straight forward comparison. In addition, similar features can be of different length in compared variables because of scaling properties. Assuming two variables have a comparable increase per time but with different amplitude ranges, then the same feature will be different lastingly. In fact, for reasonable comparison of time series, algorithms needs to compensate time warping.

    A state-of-the-art approach to compensate time warping is solving an optimisation problem for the shortest distance. In short, starting at a common point in time series, the algorithm compares each individual feature of time series A with several from time series B and searches for the best matches to minimise the distance either or maximise the similarity. Consequently, it iterates through a distance/similarity matrix comprising quantified matches for all possible combinations. Under certain conditions, this approach risks to drift constantly away from the natural matching (diagonal path) and breaks the so called Sakoe Chiba band \cite{sakoe_dynamic_1978}. For this reason, most solutions following this approach require additional constraints to force matching along the optimal path - this also holds for the proposed algorithm.

    In contrast to time relation, amplitude warping or scaling effects are less obvious but are easier to compensate for. Depending on time series' properties, the distance/similarity measure used for feature comparison allows for mitigating amplitude warping and/or scaling effects.
    This holds for average values for instance, that compensate high variability or relative changes that ignore drift effects. Nevertheless, 
    intermediate scaling changes such as temperature dependent effects can only be compensated with additional information.
    
    \item[Quantification] turns all intermediate results into a quantitative measure representing the distance or similarity between compared time series. Usually, this is a single number that represents an overall measure. It aggregates all intermediate comparison results and maps the result to a single measure. Approaches using a distance/similarity matrix for warping compensation such as the concept of the \gls{dtw} algorithm, primarily use the last element of the matrix result as final measure.

    A quantitative distance or similarity value over the whole series is mandatory for
    rough decision making. 
    With respect to a relevance metric, this value indicates if an exogenous time series comprises useful information about the target time series over all and is of relevance for target time series.
    However, in case two series only have a short sequence presenting strong similarity, it disappears in the overall weak results. As consequence, exogenous data comprising little information but with high significance for target time series will be excluded in further calculations. In particular for forecasting the excluded information might be of relevance for achieving accurate results.
    For this reason, a serious relevance metric should provide a general measure but also local relevance information.

    In compliance with the aspects discussed, for the development of the new relevance metric we treated them separately. Consequently, the new algorithm results from many studies with different combinations and variations applied on short synthetic time series comprising various characteristics. The following chapter introduces the new algorithm and discusses the solutions for all the mentioned aspects in detail.
    
\end{description}

\section{Forward Angular Relevance Metric (FARM)}
\label{sec:farm}

The proposed relevance metric can be split into three parts according to the aspects described in Section \ref{sec:asp_rel_metric}.
It results a measure for a quantified influence of an exogenous variable on a target variable. 
In fact, it returns a measure for the relevance of a time series with respect to a target series on a global and intermediate level. As a consequence it particularly focus on intermediate similarities.

Based on findings of the literature research published in \cite{christen_distance_2022}, the new algorithm relies on the concept of the \gls{amss} approach. In a similar method it equally refers to the shape of time series for distance (i.e. relevance) measure. Assuming constant sampling intervals, the algorithm relies on vectors representing changes in time series data. Yet, it differs in all three aspects from the \gls{amss} approach and extends the core concept with a new representation of the quantified measure. A key requirement for the development was robustness. This implies general applicable algorithms with a minimal parameter set.

The following subsections discuss the three aspects of the proposed algorithm starting with the distance used for compensation of warping effects. The time warping compensation, presented in second part, relies on the distance measure defined in the first aspect. Finally, the calculation of the relevance measure called quantification, relies on the time warping compensated series.

\subsection{The FARM distance}
\label{subsec:farm_dist}
For the sake of robustness, the distance or similarity measure has to provide the ability to treat with large numerical ranges for feature representations in target and exogenous time series. Relative features provide the benefit of compensating offset differences without information loss.
This advantage is also used by other approaches like \gls{amss} and \gls{ddtw} for instance.

\begin{figure}[tb]
    \centering
    \includegraphics[width=0.48\textwidth]{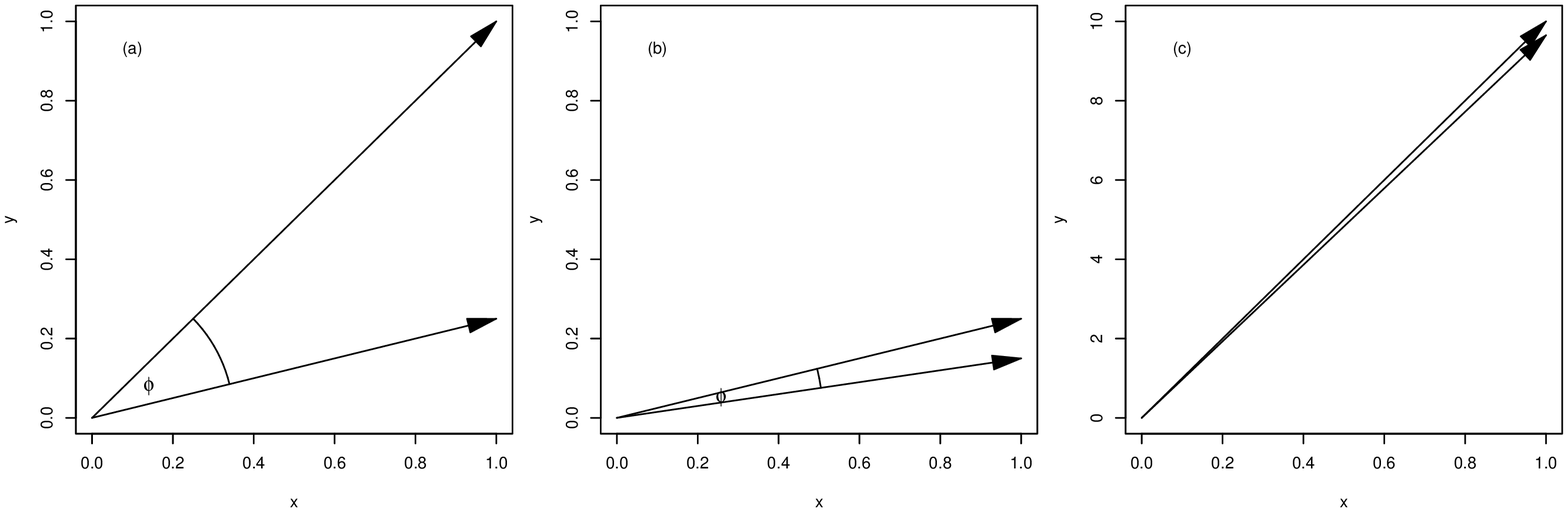}
    \caption{Interaction of vector angle and y-value. (a) Vector angle used in \gls{farm} and \gls{amss}, (b) angle for little y-value, (c) angle for large y-value.}
    \label{fig:vectors}
\end{figure}

The \gls{ddtw} extracts feature distance as linear combination of the L2 distance between the current, the previous and the next data point. In contrast, the \gls{amss} maps the time series to a vector series with a constant 
$\Delta x$ for all vectors. The $\Delta y$ equals the difference of two following data points. After that, it applies the cosine similarity of the angle $\phi$ for feature comparison, as illustrated in Figure \ref{fig:vectors}, illustration (a).

The cosine similarity provides continuous numbers from 1 for identical vector orientations (perfect match) to -1 for opposite vector orientation. Since the vectors represent the difference of two following data points with a constant time interval in between, this case only holds for a difference from $+\infty$ to $-\infty$. In contrast, perpendicular vectors have a similarity equal 0. In \gls{amss} the angle between compared vectors is limited to $\pi/2$ that results in a similarity range from 1 to 0.


In the proposed relevance metric, we use the sine distance of the angle $\phi$ similarly to the \gls{amss} approach. In this way, we still profit form difference value comparison yet with a distance instead of a similarity measure.
In contrast to the cosine similarity, the sine distance results bigger distances for small data value changes (y-value), as shown in Figure \ref{fig:vectors} illustration (b), compared to equal vector differences ($\Delta y$) for large data value changes shown in illustration (c). In other words, the distance of the feature used (i.e. the sine distance) is smaller in case the difference between data points in time series is large anyway - and vice versa. This effect makes the sine distance being a more robust feature compared to the cosine similarity.

\begin{figure}[tb]
    \centering
    \includegraphics[width=0.48\textwidth]{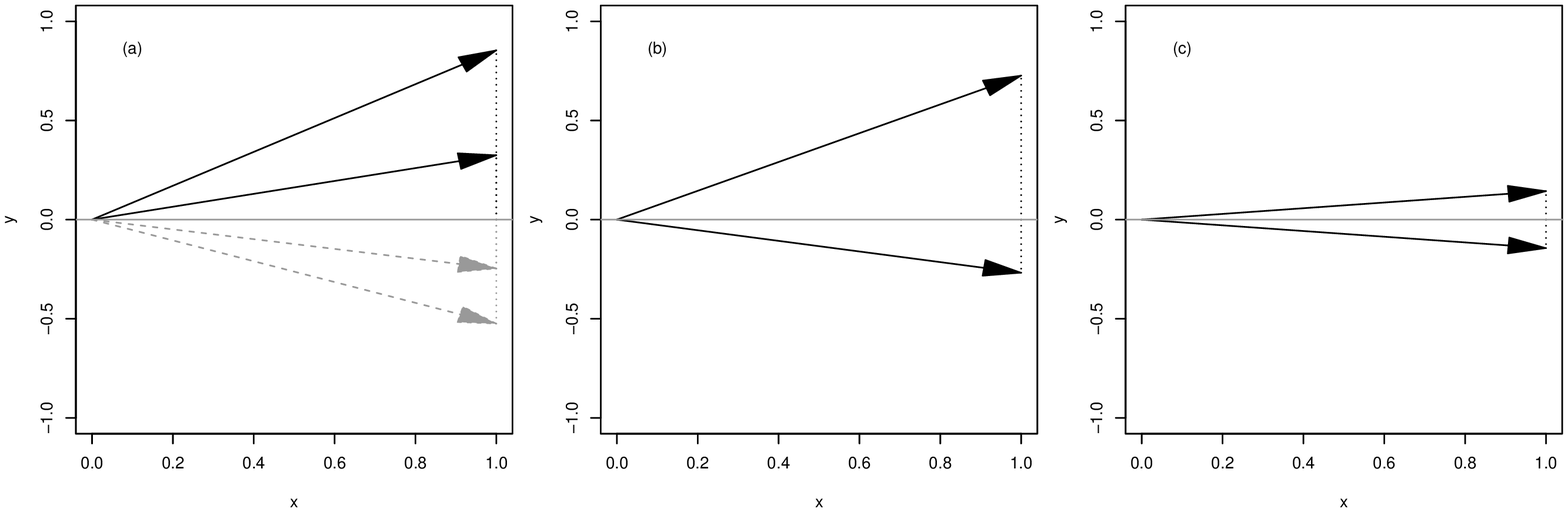}
    \caption{Cases of equal and opposite value changes ($\Delta y$).}
    \label{fig:dy_issue}
\end{figure}

The downside of the sine distance is the symmetric characteristic for angles between 0 to $\pi/2$ and $\pi/2$ to $\pi$. It results low distances if both time series have large opposite data value changes; one positive and one negative as in Figure \ref{fig:dy_issue} illustration (b). For this reason, the proposed metric constrains the sine distance to vector angles $0 <= \phi <= \pi/2$ and requires vectors to be in the same quadrant (positive or negative value changes for both series) as shown in illustration (a). 

As consequence, using the sine distance entails dropping distances measures for opposite value changes in time series. To some extent, this is coherent for large vector angles. However, small value changes in opposite direction as in Figure \ref{fig:dy_issue} illustration (c), will result in vector representations having a strong similarity. In fact, the algorithm is obliged to prefer a small distance resulting from opposite changes rather than a large distance from changes in same direction. Being conform with this requirement, the proposed solution extends the sine distance with an exponential distance for vectors of opposite value changes.


As consequence, the \gls{farm} metric combines two different measures that differently quantify the feature distance depending on the equality of the sign of data changes in compared time series.
The distinction between the two measures represents the Algorithm \ref{alg:farm_dist} in lines 4 to 8.

\begin{algorithm}[b]
    \caption{FARM distance algorithm}
    \label{alg:farm_dist}
    \textbf{Input}: $\Delta y$ for reference ($\Delta y_{ref}$) and query ($\Delta y_{qry}$) series\\
    \textbf{Output}: FARM distance feature
    \begin{algorithmic}[1] 
        \STATE Let $\phi_r=atan(\Delta y_{ref})$, $\phi_q=atan(\Delta y_{qry})$, $dist = 0$
        \IF{$sign(\phi_{ref}) == sign(\phi_{qry})$ AND $\phi_r \cdot \phi_q \neq 0$}
        \STATE $dist = sin(|\phi_{ref} - \phi_{qry}|)$
        \ELSE
        \STATE $dist = e^{|\Delta y_{ref} - \Delta y_{qry}|} -1$
        \ENDIF
        \STATE \textbf{return} $dist$
    \end{algorithmic}
\end{algorithm}

\subsection{Time series matching}
Time series rarely have an identical time base, exterior impact or feature representation. Hence, they usually show time warping effects which have to be compensated for a serious feature comparison. As consequence, the new relevance metric needs to be able to deal with warping in feature representation.

\begin{figure}[tb]
    \centering
    \includegraphics[width=0.40\textwidth]{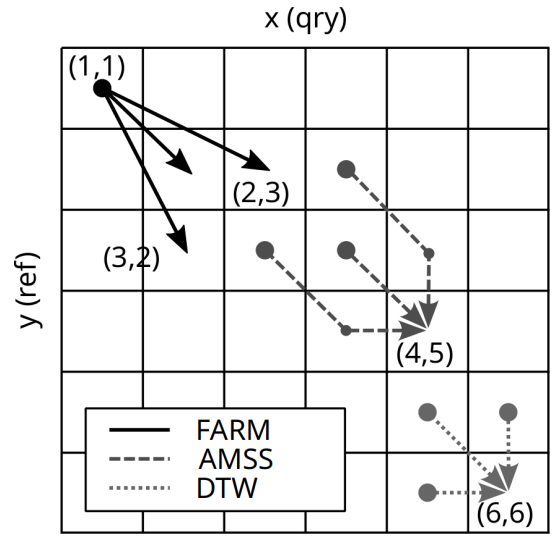}
    \caption{Search space methods to find best feature matches for warping compensation.}
    \label{fig:path_finding}
\end{figure}

The original concept for time warping compensation between time series that is proposed in \cite{berndt_using_1994} solves an optimisation problem. Given a reference (ref) and query (qry) time series, the \gls{dtw} algorithm calculates the distance between comparing features (the data values in this case) for every possible combination and creates a distance matrix as illustrated in Figure \ref{fig:path_finding}. In the same step it accumulates the least value of the previous or single shift combinations to the actual feature distance. This results in a cost matrix providing the least warping path by the lowest cost values.

In the \gls{dtw} approach, the sum of the distance value includes the distance of the current comparison and the least accumulated value of the previous (x-1, y-1), the previous in query with the current of reference time series (x-1, y) and vice versa (x, y-1). Figure \ref{fig:path_finding} shows the accumulation used in the \gls{dtw} approach for feature combination (6,6).

The \gls{amss} applies an extended version. In addition to distance combinations with previous features it also considers combinations one step back, as shown for combination (4,5).


The combination used in the \gls{amss} approach has twofold effect: for one thing, it also considers the distance one step back. Secondly, it forces a diagonal move to keep within a Sakoe-Chiba band \cite{sakoe_dynamic_1978}. These effects are relevant for improving the warping approach and are also considered in the proposed solution.

Indeed, the new \gls{farm} algorithm relies on the same concept using a distance matrix. Yet, it presents two novelties: 1. it only calculates the considered distances next to the minimal path and 2. it avoids the accumulation of previous values. In fact, the \gls{farm} algorithm selects the path with the minimal feature distance of the next straight forward reference and query combination (diagonal) and mixed feature combinations one step ahead. Completely skipping intermediate distances allows for intermediate large distances. Figure \ref{fig:path_finding} illustrates the considered distances for the next move from (1,1) to (2,2), (3,2) or (2,3). 


In this way, the new forward approach is completely different to other solutions that first build a cost matrix relying on all possible feature distances and determine the best path on it afterwards. The \gls{farm} algorithm directly determines the next step based on the minimal distance between all possible next feature combinations. As result, the next step is immediately known without any further calculation. For one thing, this reduces the calculation effort from a complexity of $O(N M)$ for a complete cost matrix to 
$O(N+M)$.
Then it also allows for a continuous approach.


A sketch of the forward alignment algorithm is given in Algorithm \ref{alg:farm}. It presents the forward approach in a while loop that allows for an iterative update of the path at each iteration. The selection of the minimal distance is shown by the less equal decisions in lines 4 and 7 that rely on the \gls{farm} distance in Algorithm \ref{alg:farm_dist}.

\begin{algorithm}[tb]
    \caption{FARM forward alignment algorithm}
    \label{alg:farm}
    \textbf{Input}: reference (ref) and query (qry) $\Delta y$ arrays\\
    \textbf{Output}: De-warping path
    \begin{algorithmic}[1] 
        \STATE Let $x=1$, $y=1$, $path=[1,1]$
        \WHILE{$y < len(ref)$ AND $x < len(qry$)}
        \STATE $next = [y+1, x+1]$
        \IF {$dist(y+2, x+1) < dist(y+1, x+1)$}
        \STATE $next = [y+2, x+1]$
        \ENDIF
        \IF {$dist(y+1, x+2) < dist(y+2, x+1)$}
        \STATE $next = [y+1, x+2]$
        \ENDIF
        \STATE $path=[path, next]$
        \STATE Update y and x
        \ENDWHILE
        \STATE \textbf{return} $path$
    \end{algorithmic}
\end{algorithm}

For the sake of robustness, the concept of the algorithm relies on a forward alignment. That means, the feature mapping is always applied ahead in time. Assuming the query time series is the leading time series (query samples are recorded before reference samples), the \gls{farm} algorithm only allows feature mapping from query ($x$) ahead to reference time series ($y$).
In fact, information must be present before it appears in target time series. This particularly holds for forecasts. As result, warping solutions can only match features in the distance matrix where $x<=y$, assuming both time (feature) series start with the same index. This implies, the number of features from reference time series is higher or equal the feature number from query time series.

\subsection{Quantified relevance measures}
Quantification transforms comparison results into a measurable value. Because of that, it is important to have features of compared time series well aligned to each other. In terms of a relevance metric, this step results a measure that quantitatively represents the relevance of an exogenous time series on a target series. 

Comparable approaches 
that also treat warping issues relying on a distance or similarity matrix usually return the last value in the matrix. By accumulating the distance and similarity values of compared features, respectively, the last value in the matrix represents the cost for fitting two time series. As consequence, this is also a reasonable measure for the general similarity of time series. Yet, in this way, 
the location of affecting features remains totally unknown. For this reason, the \gls{farm} provides two measures: intermediate (local) and a global relevance measure.

For the purpose of providing the location of relevant features, the \gls{farm} algorithm calculates a series of local correlation. In fact, it iterates over the warping compensated time series and calculates the correlation for a limited fraction covered in a window (w).
This results in a series of correlation coefficients representing local similarity between time series according to Equation \ref{eq:rel_l}. In addition, series' index describes the location of the corresponding similarity.

\begin{equation}
\label{eq:rel_l}
    \resizebox{.91\linewidth}{!}{$
            \displaystyle
            rel_l = cor(ref_{[c-w/2, c+w/2]}, qry_{[c-w/2, c+w/2]}) 
        $}
\end{equation}

The window size depends on the feature properties of time series. We used a default value of 5 with focus on fast changing data value. Yet, this parameter can vary and should be adjusted according to the properties of desired features. Balanced or slowly changing time series potentially require larger window sizes. 

Intermediate feature similarities are significant factors in measuring the global relevance (i.e. influence) of an exogenous to a target time series. As consequence, the global relevance relies on the results of the local similarity measures. According to Equation \ref{eq:rel_g}, the \gls{farm} algorithm represents the measure for the global relevance as ratio of the \gls{rms} of all local similarity measures to the correlation of full time series. In this way, the \gls{rms} more emphasises individual similarities while the full correlation in the denominator normalise the result with respect to an average similarity.

\begin{equation}
\label{eq:rel_g}
    \resizebox{.61\linewidth}{!}{$
            \displaystyle
            rel_g = \frac{\sqrt{\frac{1}{(n-w+1)} \sum_{c=1}^{n-w+1} rel_l[c]^2}}{cor(ref, qry)}
        $}
\end{equation}


Hereinafter, the pseudo code in Algorithm \ref{alg:relevance} presents the calculation of the local similarity series as well as the global relevance measure. A while loop from line 2 to 6 moves a window over the warping compensated reference (ref) and query (qry) time series and stores local results in a list. 
  

\begin{algorithm}[tb]
    \caption{FARM relevance algorithm}
    \label{alg:relevance}
    \textbf{Input}: De-warped reference (ref) and query (qry) time series\\
    \textbf{Parameter}: Window size for local relevance $(win)$\\
    \textbf{Output}: Local and global relevance
    \begin{algorithmic}[1] 
        \STATE Let $local=[]$, $global=0$, $center=win/2$
        \WHILE{center in $(series - win)$}
        \STATE $cor_{loc}=cor(ref[win], qry[win])$
        \STATE Append $local$ with $cor_{loc}$
        \STATE Increment $center$
        \ENDWHILE
        \STATE $global=\sum(local) \cdot (cor(ref, qry) \cdot len(series))^{-1}$
        \STATE \textbf{return} $[global, local]$
    \end{algorithmic}
\end{algorithm}

\section{Experiments and Constraints of FARM}
\label{sec:lim_and_compl}
This section discusses warping compensation capabilities of the \gls{farm} algorithm based on a synthetic time series used for evaluation purposes and compares the result with the \gls{amss} approach. Apart from that, it points out constraints and possible limits for the application of the \gls{farm} algorithm.

\begin{figure*}[t]
    \centering
    \includegraphics[width=\textwidth]{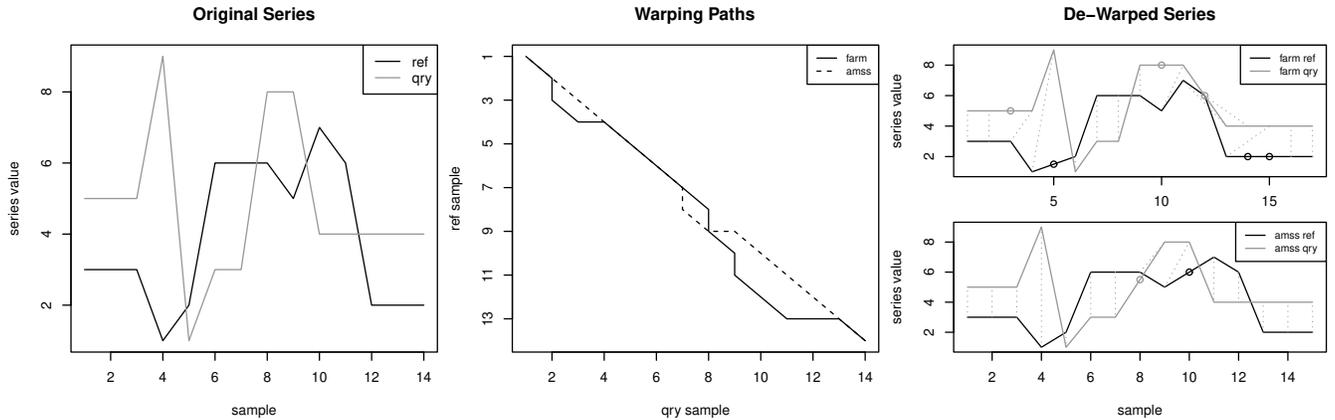}
    \caption{Warping compensation for synthetic sample time series with intermediate dissimilarities. Results for \gls{farm} and \gls{amss} approach.}
    \label{fig:ts_smple}
\end{figure*}

The design of the \gls{farm} algorithm particularly focus on similarity of remarkable features in time series. For this reason, a proper alignment of features is mandatory. Additionally, robustness is a key requirement for the algorithm. In this way, the proposed algorithm will be applicable for similarity analysis of time series from different realms. As consequence, the warping compensation performance of the algorithm has been tested in experiments on short synthetic time series presenting different characteristics that typically appear. These include positive and negative changes, different duration of equal changes as well as intermediate distortions in features, among others.

The \gls{farm} algorithm particularly relies on the methods proposed in the \gls{amss} approach. Accordingly, experiments on warping compensation are preliminary compared to results of the \gls{amss} algorithm such as in the experiment presented in Figure \ref{fig:ts_smple}.

For illustration purposes, the time series used in the presented experiment shows intermediate strong dissimilarities before or between similar features. In fact, the original query (qry) and reference (ref) time series in the left graph in Figure \ref{fig:ts_smple} show
characteristics that are not present in the compared time series. These include 
\begin{enumerate*}[label={\roman*)}]
    \item valleys at samples 4 and 9 and
    \item an overshot at sample 10 in reference series as well as
    \item a plateau in query series between samples 6 and 7.
\end{enumerate*}
In order to achieve warping compensation results 
being close to human feature assignment, these dissimilarities shall be avoided by the warping algorithm.




Predominantly, the \gls{farm} algorithm was designed for defining the relevance of exogenous variables in forecasting approaches. Based on this fact, the query time series is always in advance to the reference series. This means, features in query time series have to be in advance in time to mapped reference features. 
This fact allows for ignoring the half of the distance matrix in warping compensation. As consequence, the \gls{farm} warping path presented in the centre line diagram in Figure \ref{fig:ts_smple} can only be
within the lower left triangle and is bounded by the diagonal line where qry sample is less or equal than reference sample.


The line diagrams on the right in Figure \ref{fig:ts_smple} present the resulting adjusted time series following the resulting warping path in the centre diagram for the \gls{farm} and the \gls{amss}. 
The dashed lines indicate the original data point mapping before extension with intermediate data. Newly added data points are represented as circles and appear in the reference as well as the query time series. This results from the 
application
of the warping path on the original series - how the time warping is compensated. The \gls{farm} algorithm only inserts data points using a linear interpolation but never removes data. In this way, no information drops away. As consequence, insertions affect both time series. For comparison purposes, this method was also applied on the \gls{amss} warping result.

In this particular example, the comparison shows a more reasonable mapping (i.e. more human) for the results of the \gls{farm} than the \gls{amss} approach. This can be explained by consideration of the shifted feature distance after the current positions without adding penalty for intermediate steps. From this perspective, an increase of the ignoring step size might have an additional positive effect as it would allow more disturbing intermediate steps. 

Another aspect with significant influence on the final warping path is the treatment of opposite feature appearance. 
Depending on the recording or by chance, features may be opposite to each other: positive in one and negative in the other time series. However, in case of a common cause or a dependency, features are either strict in phase or opposed. Yet, for strict opposite features, the whole series can be inverted.
To our best knowledge, similarities with alternating alignment are rather rare.
As consequence, it is reasonable to constraint matches to positive definite features.

\begin{figure}[tb]
    \centering
    \includegraphics[width=0.48\textwidth]{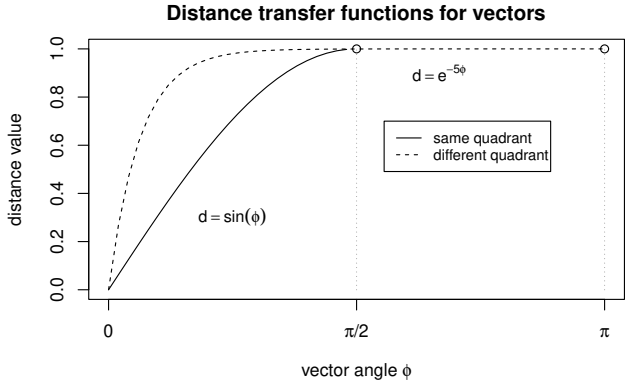}
    \caption{Distance transfer functions for vectors in same and different quadrants.}
    \label{fig:diff_tf}
\end{figure}

In fact, the \gls{farm} algorithm focus on equally aligned feature vectors as described in Subsection \ref{subsec:farm_dist}. In order to deal with small opposite data changes represented by feature vectors close to the x-axis, the algorithm applies two different distance transfer functions, shown in Figure \ref{fig:diff_tf}: (1) the sine distance for regular vectors being in the same quadrant and (2) an exponential function for opposite aligned feature vectors. However, the composed transfer function requires a multiplication by five of the full exponent in the Euler function.
The multiplication increases the slope of the function and 
ensures that the distance of the exponential function is always larger than the sine distance.
In this way, feature vectors being in the same quadrant are prioritised. 
This holds almost for the full range where $0 \leq \phi \leq \pi$ except for angles very close to $\pi$. Yet, this condition only appears for very large data value changes with a negligible small probability, in practice.

\section{Further Research}
\label{sec:further_res}

In addition to a general relevance measure, the new \gls{farm} metric also provides time-related relevance information.
This extra information allows to control in a forecast application the influence of exogenous data on the target time series time dependently. 
By applying the new metric on a real case data set, the benefit of this additional information shall be tested.

Focusing on structural changes towards distributed power generation in electrical energy domain, we intend to evaluate the new metric on
exogenous data for improving forecast of electrical power production and consumption, respectively. In fact, applied on an existing data set comprising various smart home data recordings, this implies two steps: first, the evaluation of additional information obtained by considering different exogenous variables. This will result in a reduced set of exogenous variables containing valuable information for future power consumption and/or production. Second, training an appropriate AI method to forecast the power variable with the reduced set of exogenous variables. Yet, in contrast to traditional forecast approach, we overlay the local relevance series to cancel out irrelevant information. 

The objective is to improve forecast by including selective information from exogenous data. Since the proposed \gls{farm} algorithm is applicable to time series from any domain, we will also open the evaluation to other application realms.

Finally, based on the results of real case evaluations, we will continue research in (a) interpreting distance measures with the proposed dual transfer function. (b) extending the range of feature distances considered for optimising the warping path - 
an advanced search space in distance matrix for path optimisation.
Especially for this parameter, we see a potential to further improve matching results more towards humanistic intuition. Because, an increased step size between compared features allows for having more intermediate distortions before reasonable matches. Hence, extending this parameter might allow for finding better matches. The expected downside is that it also allows for faster divergence from a diagonal path - what entails the risk of non matched features. And (c) the values used for relevance calculation. Here we see another potential by using the proposed distance value. Yet, this requires evaluation of real case examples and a full proof of the triangle inequality for the new composed distance.


\section{Conclusion}
\label{sec:conclusion}


Exploring the influence of exogenous variables on a target time series is a key for input data selection in forecast approaches. This implies the evaluation of the relevance of individual variables. Thereby,  intermediate features with strong similarity are significant for a reasonable relevance measure.

In this paper we presented a new relevance metric that more emphasise individual feature similarities in the overall similarity calculation. The new metric focus on feature comparison between time series to evaluate the relevance for forecast applications. It applies a composed distance calculation for feature comparison in time warping compensation. Based on the minimum distance, it compares shifted feature combinations in a forward approach to the next regular step. Thereby, it omits penalisation for intermediate steps having dissimilar features. In addition, the forward warping approach significantly reduces the calculation effort for evaluating the best time series alignment and warping compensation.

The time warping compensation only allows matching ahead in time. This constraint reduces the search space by half and contributes to more robustness by avoiding extensively diverging from original mapping. 

Based on more human warping results demonstrated with synthetic characteristic time series, further research preliminary focus on evaluation with real case time series and the full proof of the triangle inequality.

\bibliographystyle{named}
\bibliography{references}

\end{document}